\documentclass[11pt,a4paper]{article}
\usepackage[hyperref]{acl2019}
\usepackage{times}
\usepackage{latexsym}
\usepackage{graphicx}
\usepackage{caption}
\usepackage{subcaption}
\graphicspath{ {./images/} }

\usepackage{url}

\aclfinalcopy 

\title{MedCATTrainer: A Biomedical Free Text Annotation Interface with Active Learning and Research Use Case Specific Customisation}

\author{Thomas Searle$^{1,2}$, Zeljko Kraljevic$^1$, Rebecca Bendayan$^{1,2}$, Daniel Bean$^1$, Richard Dobson$^{1,2,3,4}$ \\
    $^1$Department of Biostatistics and Health Informatics, Institute of Psychiatry, \\ 
    Psychology and  Neuroscience, King’s College London, London, U.K.\\
    $^2$NIHR Biomedical Research Centre at South London and Maudsley \\ NHS Foundation Trust and King’s College London, London, U.K. \\
    \{$^3$Institute of Health Informatics$\vert$$^4$Health Data Research UK London\},\\University College London, 222 Euston Road, London NW1 2DA, U.K.\\
    \{firstname.lastname\}@kcl.ac.uk}
\date{}

\begin{document}
\maketitle
\begin{abstract}
We present MedCATTrainer\footnote{https://www.youtube.com/watch?v=lM914DQjvSo} an interface for building, improving and customising a given Named Entity Recognition and Linking (NER+L) model for biomedical domain text. NER+L is often used as a first step in deriving value from clinical text. Collecting labelled data for training models is difficult due to the need for specialist domain knowledge. MedCATTrainer offers an interactive web-interface to inspect and improve recognised entities from an underlying NER+L model via active learning. Secondary use of data for clinical research often has task and context specific criteria. MedCATTrainer provides a further interface to define and collect supervised learning training data for researcher specific use cases. Initial results suggest our approach allows for efficient and accurate collection of research use case specific training data.
\end{abstract}

\section{Introduction}
We present a flexible web-based open-source use-case configurable interface and workflow for biomedical text concept annotation - MedCATTrainer\footnote{https://github.com/CogStack/MedCATweb}. 

 \citet{Murdoch2013-ld} estimates that 80\% of biomedical data is stored in unstructured text such as Electronic health records (EHRs). Although EHRs have seen widespread global adoption, effective secondary use of the data remains difficult \cite{Elkin2010-cj}. However, significant progress has been made on agreement and usage of standardised terminologies such as the Systematized Nomenclature of Medical Clinical Terms (SNOMED-CT) \cite{Stearns2001-qf} and the Unified Medical Language System (UMLS)\cite{Bodenreider2004-ir}. Annotating EHR text with these concept databases is often seen as a first step in delivering data driven applications such as precision medicine, clinical decision support or real time disease surveillance \cite{Assale2019-aq}.

EHR text annotation is challenging due to the use of domain specific terms, abbreviations, misspellings and terseness. Text can also be `copy-pasted' from prior notes, structured tables entered into unstructured form, content with varying temporality and scanned images of physical documents \cite{Botsis2010-hx}. Annotation is further complicated as researchers have task and context specific parameters. For example, whether family history or suspected diagnoses are considered relevant to the task.

MedCAT\footnote{https://github.com/CogStack/MedCAT}, manuscript in preparation \cite{kraljevic_zeljko_2019_3265442}, is a \textbf{Med}ical \textbf{C}oncept \textbf{A}nnotation \textbf{T}ool that uses unsupervised machine learning to recognise and link medical concepts with clinical terminologies such as UMLS. MedCAT, like similar tools, uses a concept database to find and link concept mentions inside of biomedical documents. In addition it has disambiguation, spell-checking and the option for supervised learning for improved disambiguation.

We introduce a novel web based application that supplements usage of biomedical NER+L models, such as MedCAT. Our contributions are as follows:
\begin{enumerate}
  \item An interface that wraps around a NER+L system that enables users to inspect the identified concepts from free text, and add missing concepts to the NER+L model. This interface aligns with MedCAT, but could also be used with other models that have similar capabilities.
  \item An interface for active learning, enabling users to provide minimal training data to assist in improving and correcting the NER+L. This interface requires that the NER+L system supports active learning.
  \item A further interface for configurable use case specific annotation of identified concepts. Allowing for collection of research question specific training data. For example, annotating specific temporal features of a concept.
\end{enumerate}

\section{Related Work}
Outside of the biomedical domain general purpose annotation interfaces have been developed for most popular NLP tasks such as NER, NEL, relation extraction, entity normalisation, dependency parsing, chunking etc. Popular choices include open-source tools such as BRAT \cite{Stenetorp2012-rr} that also allow for managing the distribution, monitoring and collection of annotated corpora. General purpose tools that include active learning include the open-source DUALIST \cite{Settles2011-uf} and the commercial product Prodigy\footnote{https://explosion.ai/blog/prodigy-annotation-tool-active-learning}. Although these tools are mature and offer advanced features they can be complex to setup and do not offer integration with existing biomedical domain NER+L systems.

Prior work on biomedical NER+L includes MetaMAP \cite{Aronson2001-kr} and CTakes \cite{Savova2010-bv}. Both have provided interfaces to inspect recognised entities but they have not provided means to correct and amend concepts or specify further annotations for specific research questions.

Another tool for biomedical NER+L, SemEHR \citet{Wu2018-yw}, offers features to add custom pre and post processing steps and research specific use cases, but does not directly improve the NER+L model via an interface. Instead it treats the provided NER+L model as a black-box model.

\section{MedCATTrainer}
MedCATTrainer is a web-based interface for inspecting, adding and correcting biomedical NER+L models through active learning. An additional interface allow research specific annotations to be defined and collected for training of supervised learning models.

The interfaces are built with Vue.js\footnote{https://vuejs.org/} for the front-end and the python\footnote{https://www.python.org/} web framework Django\footnote{https://www.djangoproject.com/} for the web API and integration with NER+L models such as MedCAT. We use the Django admin features to allow administrators to configure research question specific supervised learning tasks.

MedCATTrainer is deployed via a Docker\footnote{https://www.docker.com} container. This ensures users can build, deploy and run MedCATTrainer cross-platform without lengthy build and run processes, advanced infrastructure knowledge or root access to systems. This is especially important in health informatics as information technology (IT) infrastructure is often restrictive.  MedCATTrainer allows researchers to build on top of existing biomedical domain ontologies, such as UMLS, for two use cases. Firstly, improving the underlying NER+L model (e.g. MedCAT) by adding synonyms, abbreviations, multi-token concepts and misspellings directly from the interface. Secondly, by allowing research use case specific annotations to be defined and collected for training of supervised learning models.

\subsection{Concept Inspection and Addition}

\begin{figure*}
    \begin{subfigure}[b]{0.5\textwidth}
        \centering
        \includegraphics[scale=0.22]{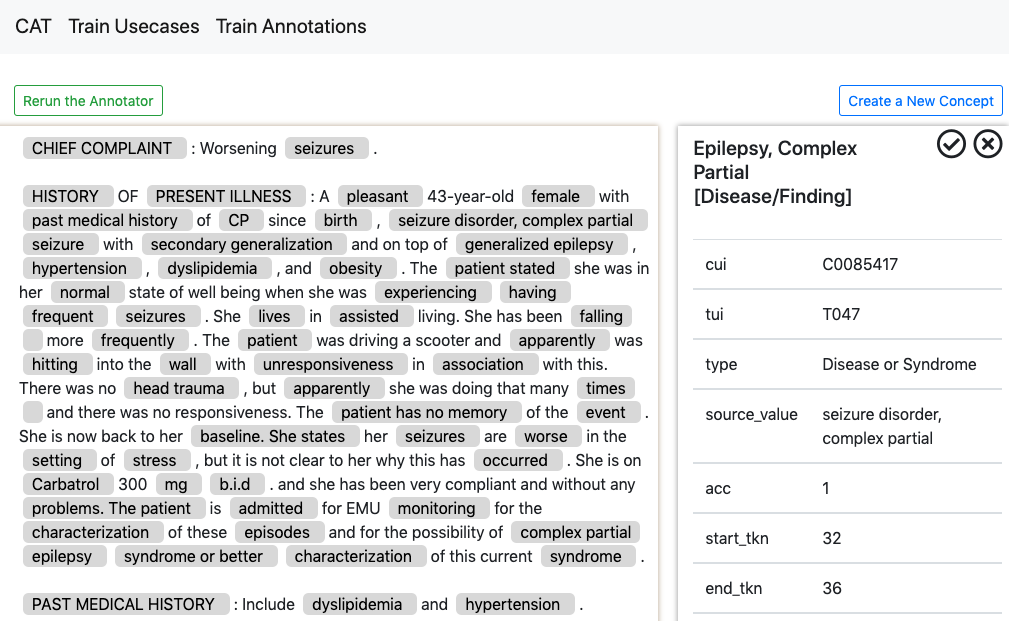}
        \caption{The MedCATTrainer interface for viewing identified concepts by the underlying NER+L model of a publicly available\footnote{https://bit.ly/2RLcdJx} neurological consultation summary showing the concept metadata and active learning feedback input controls.}
        \label{fig:screen1}
    \end{subfigure}
    \begin{subfigure}[b]{0.5\textwidth}
        \centering
        \includegraphics[scale=0.21]{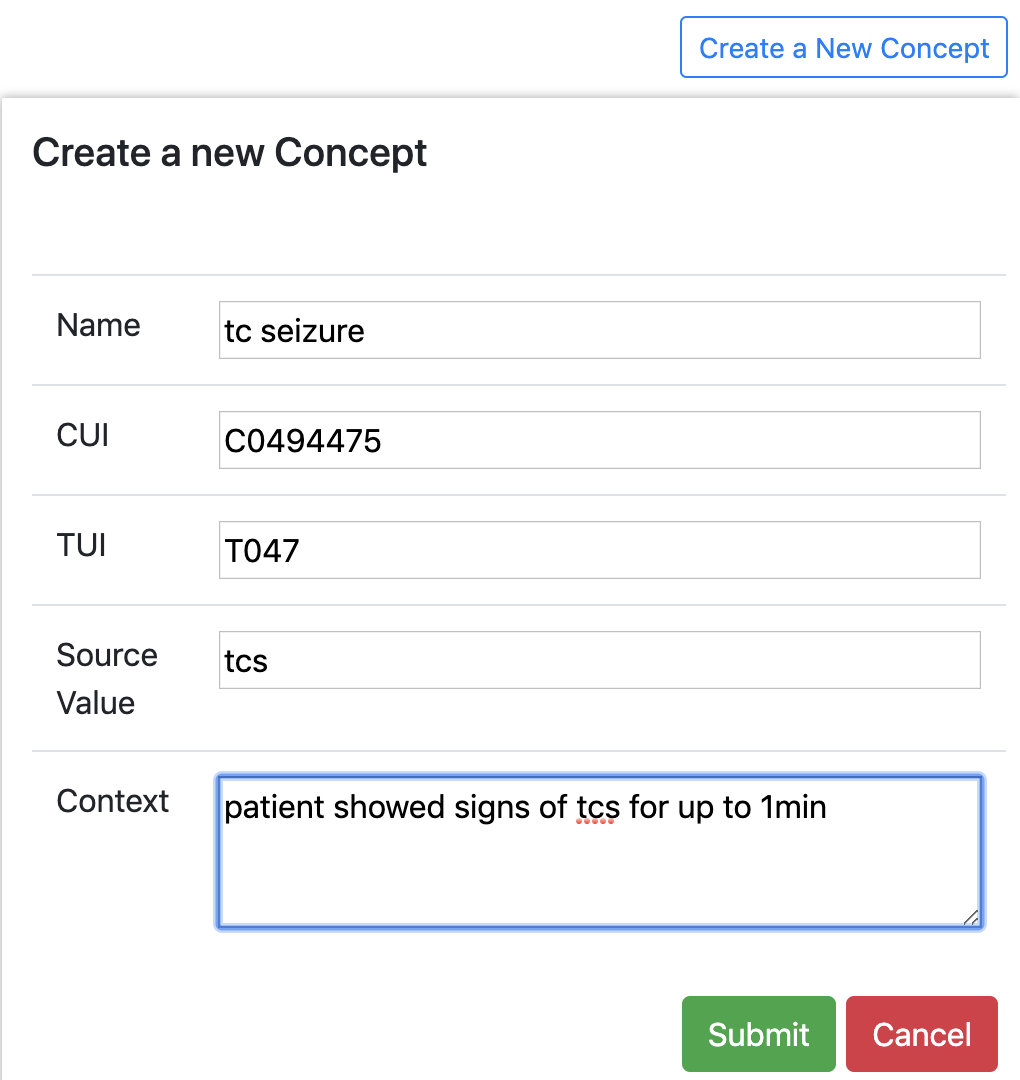}
        \caption{Side panel for the addition of new concepts.}
        \label{fig:add_concept}
    \end{subfigure}
\end{figure*}

Figure \ref{fig:screen1} shows the "Train Annotations" interface. Users can inspect and correct the concepts identified by the underlying NER+L model. Entities that have not been recognised can also be added to the NER+L model concept database. This allows researchers to test the learnt entity recognition/linking capabilities of the model whilst tailoring it to recognise sub-domain specific lexicon. This can include abbreviations or misspellings common to specific corpora. Figure \ref{fig:add_concept} shows the form entry to add new concepts to the underlying concept database. Texts that should be identified as semantically equivalent concepts can be added under the same Concept Unique Identifier, and further synonyms can also be added explicitly. Advanced NER+L tools (e.g. MedCAT) learn from the contextual embeddings of words to disambiguate future occurrences. MedCATTrainer provides a text-box for entering the context, surrounding tokens either side of the concept, to assist with concept disambiguation.

\subsection{Active Learning}
Annotating biomedical domain text for NER+L requires expert knowledge and therefore cannot be easily crowd sourced. Active learning is a common approach to provide a minimal set of high value training examples for manual annotation. Examples are valued with respect to expected improvement in classification performance once labelled and the model retrained \cite{Settles2009-br}.

We use a simple strategy of certainty based selective sampling \cite{Lewis1994-om} to display low confidence examples. Concretely, given a trained model $ \mathbf{M} $, and the total set of annotations predicted on a new document $d$ by model $\mathbf{M}$ is $ \mathbf{L} = \{l_1, l_2, \ldots l_n\} $ where the model labelled the document with $n$ annotations. An annotation $l_i$ has an associated confidence $c_{l_i}$ probability in the annotation. An annotation manager defines $ \delta $, a confidence cutoff score. The set of annotations $\mathbf{A}$ shown to an annotator is therefore $\Phi(\mathbf{L})$ where $\Phi(l_i) = c_{l_i} > \delta $.

Each human annotator is instructed to review each identified concept and provide feedback on correctness. Feedback is provided through the action of clicking either the `tick' for correct or `cross' for incorrect as shown in top right corner of the right side panel in Figure \ref{fig:screen1}.

If an identified concept is incorrect human annotators are asked to provide 'incorrect' feedback, rerun the NER+L model (top left `Rerun the Annotator'), and then confirm if the misidentified concept has been corrected. More feedback can be provided if needed. Our pilot test users found this quickly resulted in the correctly identified and linked concept as text spans often only have one or two alternative concepts.

\subsection{Clinical Research Question Specific Annotation}
It would be infeasible to have a clinical terminology to define every possible contextual representation of a concept. For example, disambiguation of `seizure' for a symptom of epilepsy and `first seizure clinic' for a clinic that provides epilepsy care or 'history of seizures' for a historical case of epilepsy.

\begin{figure*}
    \centering
    \includegraphics[scale=0.25]{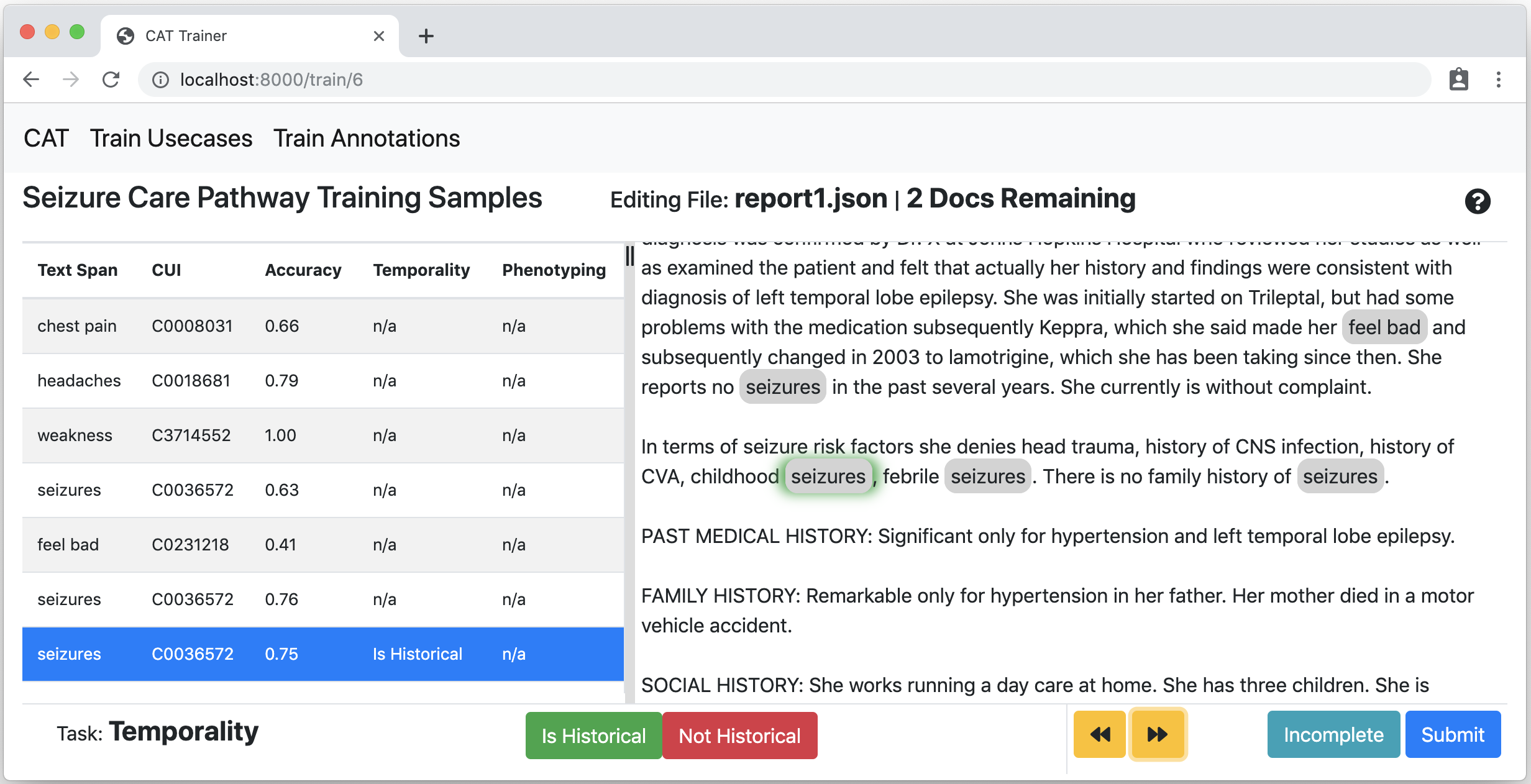}
    \caption{Task and context specific annotation interface configured for `Temporality' and `Phenotype' tasks}
    \label{fig:use_case}
\end{figure*}

Our second interface solves this problem by allowing clinical researchers to define use case orientated tasks and associated annotations for previously identified and linked concepts. Custom classifiers are then trained and layered over the existing NER+L model for context specific concept disambiguation. An example configured screen for 'Temporality' and 'Phenotyping' tasks for an ongoing clinical research project is shown in Figure \ref{fig:use_case} - using replacement publicly available data. The top bar lists the overall task name followed by the number of documents to be annotated. The top right corner opens the current task help document, listing annotation guidelines for this use-case. 

The left panel itemises each text span, the associated Concept Unique Identifier (CUI) - that the NER+L model has identified and linked with the text, and the current value of each task specific annotation. N/A indicates the task has not been completed for that span. Users can choose any order of the text spans to annotate. The currently selected text span is highlighted in the table and within the central text area showing the entirety of the document. Clinical notes can be long in length. Clicking a text span from the sidebar scrolls the central text area to the corresponding span assisting human annotators in locating the to annotate. The text area also highlights each spans current annotated value for the current task. 

The bottom bottom bar lists the current task and the possible annotation values. Figure \ref{fig:use_case} shows the `Temporality' task and the associated annotation values `Is Historical' and `Not Historical`. These values are defined in regard to this seizure care pathway use case and are defined as any currently experienced mention of symptoms experienced in this clinical encounter. Task values are configurable via the admin interface.

The bottom right corner provides navigation between text spans and tasks via the arrow buttons. Navigating between spans highlights the current span to be annotated in the main left sidebar and auto scrolls to the next span in the main text area. The navigation controls here, the sidebar and the main text area allow human annotators to complete the task in any order they are comfortable.

The `Incomplete' button marks the current document to be revisited at a later date. Samples are marked incomplete if the NER+L model has misidentified the concept or there is a genuine ambiguity. The `Submit' button marks the document as complete. Both actions store and retrieve the next document if there is one available. If there are no more files to annotate a dialog prompts the user to return to the home screen.
 
Corpora are currently directly uploaded via a use case management screen. Future deployments will directly ingest documents via an elasticsearch\footnote{https://www.elastic.co/} connector to hospital EHR deployments of CogStack \cite{Jackson2018-an} an EHR ingestion, transformation and search service deployed at King's College Hospital (KCH) and South London and Maudsley(SLaM) NHS Foundation Trusts, UK.

\section{Results}

We ran an initial small scale pilot experiment to test the suitability of our use case specific tool to quickly and accurately collect training data labelling the temporal features of seizure symptoms. This is similar to the task shown in Figure \ref{fig:use_case}. We used MIMIC3 \cite{Johnson2016-np}, a de-identified publicly available database of ICU admission data that includes observations, consultation and discharge summary reports. We randomly sampled 127 discharge summaries that contained one or more token occurrences that match the regular expression `seizure$\vert$seizre$\vert$seizur$\vert$siezure', where $\vert$ is an OR operator between the text tested to be present. We intentionally rely on a rule-based NER mode (i.e. the regex) here to demonstrate our tools flexibility to use possible alternatives to MedCAT if desired.

We asked 2 human non-clinical annotators to label temporal features of each occurrence in  relation to a 'present', i.e. `chief complaint: seizure' or `historical', i.e. `family history of seizures', mention of the term. Both took approximately 35 minutes to review all 127 documents. We achieve an percent agreement of 89\% and a Cohen's Kappa $\kappa = 0.695$, Table \ref{tab:use_case_testing}. Both annotators marked some records as incomplete as they either mostly referred to non symptomatic mentions of seizure, i.e. `anti-seizure meds prophylaxis' or they prevention of future seizures. This resulted in each rater having differing total documents `submitted' as there are some document with mixes of the above occurrences. We took the intersection of submitted documents to compute the final agreement scores.

\begin{table}
    \centering
    \begin{tabular}{|c|c|c|c|c|}
    \hline
 	&   R1*	& R2* & R1 & R2 \\
        \hline
    \# Documents & 107 & 117 & 100 & 100 \\     
    \# Concepts & 351 & 344 & 317 & 317 \\
    \# Historical & 67 & 80 & 79 & 65 \\
    \# Not Historical & 276 & 264 & 238  & 252\\
    \hline
    \end{tabular}
    \caption{Total labelled `seizure' symptom concepts and for each human annotator (R1, R2) for the `temporality' task of labelling concepts that have occurred the past relative to the hospital episode. * indicates raw numbers before taking into account the intersection of notes between annotators}
    \label{tab:use_case_testing}
\end{table}

Scikit-learn\footnote{https://scikit-learn.org/stable/index.html} was used to fit the following model. We took took each occurrence where both annotators agreed, took a random 70/30 train test split, used a TF-IDF vectoriser with the default english stop-words list and ran a grid search across TF-IDF and random forest classifier parameters with a 3 fold cross validation. We found the best fitting parameters included:  TF-IDF features 500 (range:500, 1000, 10000), maximum number trees of 100 range(100, 300, 500, 1000) and maximum tree depth 20 (range: 5, 20, 50, 75). We achieve an accuracy of this binary classification task of 92\% and f1 score .79.

\section{Discussion and Future Work}
From our labelling exercise we demonstrate the speed and accuracy of our configurable use case specific interface. Strong scores across \% agreement, Cohen's Kappa and trained model accuracy indicate good agreement between annotators, interpretations of the task and reasonable signal captured even with this small data set. Although, it is likely the model is over-fitting due to the size of the data set. Given the prior  experiment - across two raters - gathering enough accurate data to, for example, fine-tune a pretrained language model based classifier would be of the order of hours of manual labelling for approx 2k samples. We see this rapid labelling ability as a key strength of our interface. 

We foresee that trained classifiers will likely generalise to additional research questions. For example language used to express temporality of seizures is likely to be similar to temporality of stroke or myocardial infarction.

Generally, training models across use cases will likely capture shared semantics. This suggests particular use cases would require less examples to train as annotated data or the model itself could be reused, therefore  jump-starting clinical research. If a model is not performing for a new use case, further data could be collected to fine tune the model to a specific task, context or sub-domain corpora.

Clinically, domain experts in the neurology department of KCH, with varying levels of expertise (medical student to practising consultant) are scheduled to participate in the use case shown in Figure \ref{fig:use_case} in the coming months.

Our initial testing, not shown above due to space, of the active learning approach for improving the bound NER+L model suggests we can improve performance with minimal training data.
\section{Conclusions}
We have presented a lightweight, flexible, web-based, open-source annotation interface for biomedical domain text. MedCATTrainer is integrated with a biomedical NER+L model and allows for addition of missing concepts, improvements to the underlying NER+L model through active learning, and a configurable interface for clinical researchers to define annotations specific for their research questions. Preliminary results show promise for our interface and our approach to biomedical NER+L, which is often seen as a first step in deriving value from data sources such as electronic health records.

\section*{Acknowledgments} 
DMB is funded by a UKRI Innovation Fellowship as part of Health Data Research UK (MR/S00310X/1). RB is funded in part by grant MR/R016372/1 for the King’s College London MRC Skills Development Fellowship programme funded by the UK Medical Research Council (MRC) and by grant IS-BRC-1215-20018 for the National Institute for Health Research (NIHR) Biomedical Research Centre at South London and Maudsley NHS Foundation Trust and King’s College London. RD's work is supported by 1.National Institute for Health Research (NIHR) Biomedical Research Centre at South London and Maudsley NHS Foundation Trust and King’s College London. 2. Health Data Research UK, which is funded by the UK Medical Research Council, Engineering and Physical Sciences Research Council, Economic and Social Research Council, Department of Health and Social Care (England), Chief Scientist Office of the Scottish Government Health and Social Care Directorates, Health and Social Care Research and Development Division (Welsh Government), Public Health Agency (Northern Ireland), British Heart Foundation and Wellcome Trust. 3. The National Institute for Health Research University College London Hospitals Biomedical Research Centre. This paper represents independent research part funded by the National Institute for Health Research (NIHR) Biomedical Research Centre at South London and Maudsley NHS Foundation Trust and King’s College London. The views expressed are those of the author(s) and not necessarily those of the NHS, MRC, NIHR or the Department of Health and Social Care.
\bibliography{acl2019}
\bibliographystyle{acl_natbib}

\end{document}